\documentclass[a4paper]{article}

\usepackage[english]{babel}
\usepackage[utf8x]{inputenc}
\usepackage[T1]{fontenc}

\usepackage[a4paper,top=3cm,bottom=2cm,left=3cm,right=3cm,marginparwidth=1.75cm]{geometry}

\usepackage{amsmath}
\usepackage{graphicx}
\usepackage{tabularx}
\usepackage[colorinlistoftodos]{todonotes}
\usepackage{setspace}
\usepackage{siunitx}
\usepackage{breakcites}
\usepackage{natbib}

\date{}
\title{\bf Sequencing single-stranded libraries on the Illumina NextSeq 500 platform}

\author{
\large
\textsc{Johanna L.A. Paijmans\textsuperscript{1,*}, Sina Baleka\textsuperscript{1}, Kirstin Henneberger\textsuperscript{1},}\\
\textsc{Ulrike H. Taron\textsuperscript{1}, Alexandra Trinks\textsuperscript{2}, Michael V. Westbury\textsuperscript{1}, }\\
\textsc{Axel Barlow\textsuperscript{1,*}}\\[2mm] % Your name
\normalsize 
1. Institute for Biochemistry and Biology, University of Potsdam, Potsdam, Germany \\ 
\normalsize 2. IRI Life Sciences, Humboldt University Berlin, Berlin, Germany \\
\small * Corresponding Authors: \href{mailto:paijmans.jla@gmail.com}{paijmans.jla@gmail.com}, \href{mailto:axel.barlow.ab@gmail.com}{axel.barlow.ab@gmail.com} % Your email address
}

\begin{document}
\maketitle

\onehalfspacing

\section*{Introduction}

In recent years, a major contribution to the field of palaeogenomics has been the development of a novel Illumina\textsuperscript{\textregistered} sequencing library preparation protocol based on single-stranded DNA \citep{gansauge_single-stranded_2013,gansauge_single-stranded_2017,meyer_high-coverage_2012-1}. This innovative method is particularly suited for the recovery of extremely short DNA fragments typically found in ancient samples, and is associated with greatly increased conversion efficiency compared to alternative methods \citep[e.g.][]{barlow_massive_2016}). Although the method was developed specifically for ancient DNA, it has also found application in other fields such as medicine (e.g. prenatal testing from amniotic fluid \citep{karlsson_amplification-free_2015} and transplantation medicine \citep{burnham_single-stranded_2016}) and analysis of preserved tissue samples \citep[e.g. formalin-fixed paraffin embedded cells;][]{stiller_single-strand_2016}.

\begin{figure}
\centering
\includegraphics[width=\textwidth]{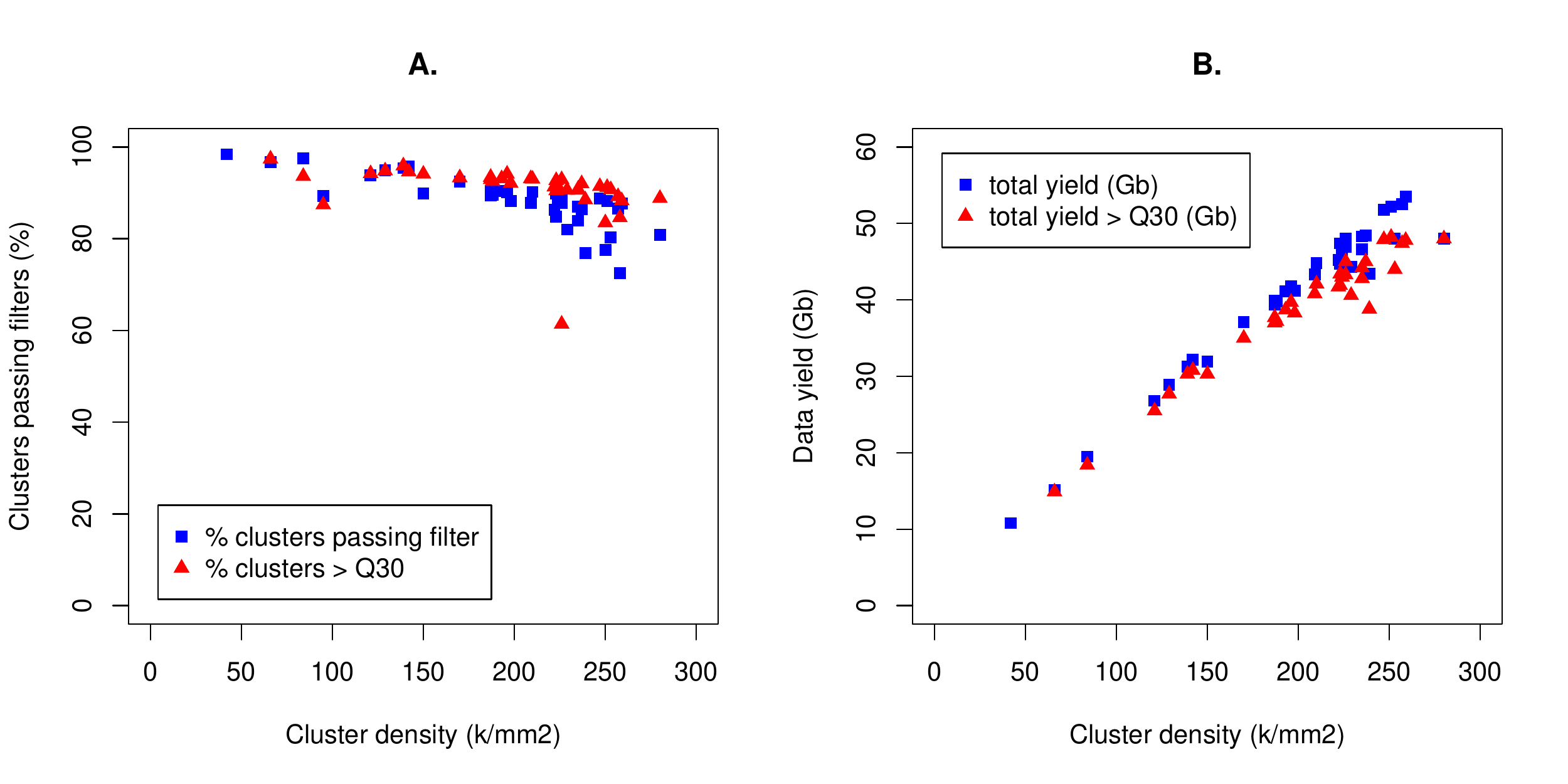}
\caption{\label{figure:graph} A) Percentage of clusters passing the filters (y-axis) versus the cluster density (x-axis) for single-stranded libraries with a 2.2 pM loading quantity. For the majority of runs, cluster densities up to 250 k/mm2 still yield >80\% passing both the chastity filter and the q30 filters. B) Data yield (y-axis) versus cluster density (only taking single-stranded library sequencing runs on 75 cycle high-output kit into account). Data yield increases more or less linearly with increasing cluster density, with only slight drop off observed around 200-250 k/mm2.}
\end{figure}

Single-stranded libraries have specific sequencing requirements. They require a custom read 1 sequencing primer. In addition, the use of dual index barcodes may also require a custom index read 2 sequencing primer on some Illumina platforms. Finally, short library molecules associated with ancient DNA templates are expected to behave differently during the annealing to the flow cell and subsequent cluster generation (bridge PCR), requiring optimisation of loading amounts to obtain optimal and consistent cluster densities. 

For the past 3 years, we have carried out sequencing of single-stranded libraries on the Illumina NextSeq 500 sequencing platform at the Institute for Biochemistry and Biology, University of Potsdam. Achieving consistent, high quality data outputs required considerable optimisation and the design of a novel index 2 read sequencing primer (Table \ref{table:primers}). We found that the short library molecules typically generated from ancient DNA templates do not produce clusters on the flow cell as efficiently as modern, standard double-stranded libraries. In our experience, single-stranded libraries benefit from an increased loading quantity for sequencing (2.2 pM) compared with double-stranded libraries (2.0 pM). Note that both these loading quantities are higher than that typically recommended by Illumina, which is 1.8 pM (NextSeq System Denature and Dilute Libraries Guide, Document \#15048776, January 2016). Furthermore, we have found that for single-stranded libraries, cluster densities can be raised substantially above that recommended by Illumina (recommended 129 - 165 k/mm2; https://www.illumina.com/systems/sequencing-platforms/nextseq/specifications.html), greatly increasing sequencing yield with little reduction in data quality (\string>200 k/mm2; Fig. \ref{figure:graph}). 

We report our optimisations here. This document may be useful for other researchers wishing to sequence single-stranded libraries on the NextSeq 500 platform. It does not replace the excellent documentation provided by Illumina (links provided below), but rather serves as additional information specific to single-stranded libraries. The original papers describing the library protocol should also be studied in detail, and complement the information presented here. 

\begin{itemize}
\item Single-stranded library protocol \citep{gansauge_single-stranded_2013}
\item\href{https://support.illumina.com/content/dam/illumina-support/documents/documentation/system_documentation/nextseq/nextseq-custom-primers-guide-15057456-01.pdf}{Custom primer guide} 
\item\href{https://support.illumina.com/content/dam/illumina-support/documents/documentation/system_documentation/nextseq/nextseq-denature-dilute-libraries-guide-15048776-02.pdf}{Denature and diluting libraries guide}
\item\href{https://support.illumina.com/content/dam/illumina-support/documents/documentation/system_documentation/nextseq/nextseq-500-system-guide-15046563-02.pdf}{Full NextSeq systems guide}
\end{itemize}

\begin{table}[b]
\centering
\begin{tabularx}{1.03\textwidth}{l l p{4.5cm} l}
\hline
\textbf{Sequencing read} & \textbf{Primer name} & \textbf{Sequence (5'-3')} & \textbf{Reference} \\
\hline \hline
Read 1 & CL72 & ACA CTC TTT CCC TAC ACG ACG CTC TTC C & \citet{gansauge_single-stranded_2013} \\
Index Read 2 & Gesaffelstein* & GGA AGA GCG TCG TGT AGG GAA AGA GTG T & This document \\
\hline
\end{tabularx}
\caption{\label{table:primers} Custom oligos required for sequencing dual indexed single-stranded libraries on the NextSeq 500. Oligos should be synthesized on a 0.05-\si{\micro}mole scale and purified by reverse-phase HPLC. Stock concentrations should be 100 \si{\micro}M. See also Gansauge \& Meyer 2013. * = the accepted abbreviation for this oligo, as may be written on a microcentrifuge tube, is ``G'stein''}
\end{table}

Finally, these procedures are what works well in our laboratory. We provide no guarantees for the success or failure of sequencing runs performed according to these recommendations. For any further discussion readers are also welcome to contact the corresponding authors by email. We are interested in combining our experiences with those of other people with a view to further optimise sequencing single-stranded libraries on the NextSeq platform.

\section*{Library pooling and quantification using Qubit and Tapestation}

\begin{figure}
\centering
\includegraphics[width=0.6\textwidth]{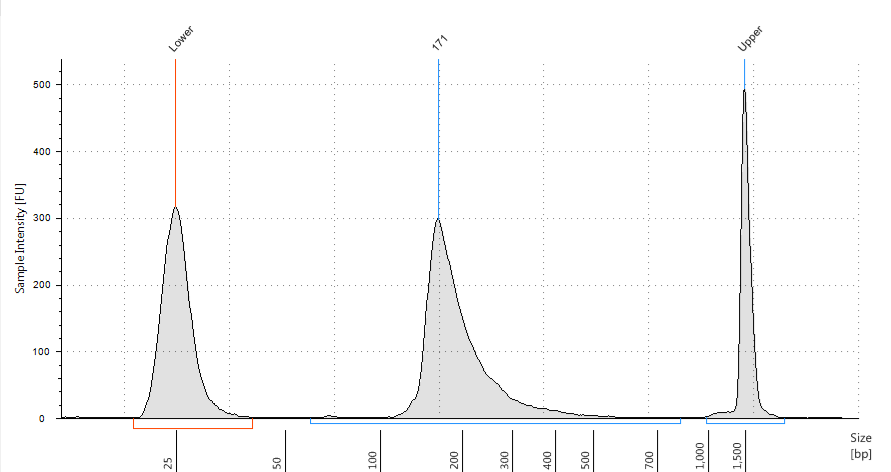}
\caption{\label{figure:tapestation} Tapestation Electropherogram report from a typical single-stranded library: approximately 5 nM, D1000 High-Sensitivity ScreenTape}
\end{figure}

\begin{enumerate}
\item The following recommendations for measuring the concentration and the modal fragment size allow the molarity of the sequencing pool to be calculate. The target concentration is 4nM, but reliable sequencing can be performed using lower concentrations (Table \ref{table:concs}). We find that a library with starting concentration less than 2.5 nM gives inconsistent results. Ideally, libraries comprising the pool should be unimodal and free of any obvious heteroduplexes or adapter/primer artefacts (Fig. \ref{figure:tapestation}). These factors will complicate library quantification and can lead to sub-optimal cluster densities.
\item We recommend using a dedicated pipette set that is regularly calibrated, as small inaccuracies can lead to unreliable and inconsistent sequencing results.
\item Measure the concentration with the Qubit HS DNA Assay (Qubit 2.0), using at least 10 \si{\micro}l of the sequencing library. Using less than 10 \si{\micro}l for quantification leads to unreliable results and thus unreliable cluster density.
\item We recommend using a 1:2 dilution of the Standard \#2 from the Qubit BR Assay as quantification control; this should be roughly 50 ng/\si{\micro}l and will allow for a verification of the quantification (use a different stock than the one used in the measurement).
\item Run sequencing pool on a D1000 HS ScreenTape (Agilent Tapestation 2200; Fig. \ref{figure:tapestation}) to estimate the modal fragment length
\item Calculate the molarity based on the modal fragment length and measured concentration. Although the modal length will not represent the true average as the fragment length distribution for single-stranded libraries is typically skewed, we find this to yield the most consistent method of quantification. We find this approach is able to produce consistent and predictable cluster densities without the need for qPCR.
\end{enumerate}

\begin{table}[t!]
\centering
\begin{tabularx}{0.85\textwidth}{p{4cm} c c}
\hline
\textbf{Starting concentration \ of library} & \textbf{Volume Library/NaOH} & \textbf{Volume HT1 Buffer} \\ \hline \hline
4 & 5 & 985 \\
3.9 & 5.13 & 984.62 \\
3.8 & 5.26 & 984.21 \\
3.7 & 5.41 & 983.78 \\
3.6 & 5.56 & 983.33 \\
3.5 & 5.71 & 982.86 \\
3.4 & 5.88 & 982.35 \\
3.3 & 6.06 & 981.82 \\
3.2 & 6.25 & 981.25 \\
3.1 & 6.45 & 980.65 \\
3 & 6.67 & 980 \\
2.9 & 6.9 & 979.31 \\
2.8 & 7.14 & 978.57 \\
2.7 & 7.41 & 977.78 \\
2.6 & 7.69 & 976.92 \\
2.5 & 8 & 976 \\ \hline
\end{tabularx}
\caption{\label{table:concs} Volumes of libraries for denaturing and dilution stages in the \href{https://support.illumina.com/content/dam/illumina-support/documents/documentation/system_documentation/nextseq/nextseq-denature-dilute-libraries-guide-15048776-02.pdf}{NextSeq sequencing protocol}. The volumes follow the formula 20\string^1/nM.}
\end{table}

\newpage
\mbox{~}

\bibliographystyle{apalike}
\bibliography{MyLibrary}

\end{document}